\documentclass[fleqn,usenatbib]{mnras}
\usepackage{newtxtext,newtxmath}

\usepackage[T1]{fontenc}
\usepackage{ae,aecompl}
\usepackage{graphicx}	
\usepackage{amsmath}	
\usepackage{amssymb}	
\usepackage{stmaryrd}
\usepackage{array,multirow}
\usepackage{bm}
\usepackage{subfig}
\usepackage{inputenc}
\usepackage{subfig}

\title[Alignment of Radio Galaxy Axes]{Alignment of Radio Galaxy Axes using FIRST Catalogue}

\author[M. Panwar, Prabhakar, P. K. Sandhu, Y. Wadadekar and P. Jain]{
Mohit Panwar,$^{1}$\thanks{email: mohitpan@iitk.ac.in}
Prabhakar,$^{1}$\thanks {email: prabhak@iitk.ac.in}
Pritpal Kaur Sandhu,$^{1}$\thanks {email: physicspritpal04@gmail.com}
\newauthor{Yogesh Wadadekar,$^{2}$\thanks {email:  yogesh@ncra.tifr.res.in}
Pankaj Jain $^{1}$\thanks {email:  pkjain@iitk.ac.in}}
\\
$^1$Physics Department, Indian Institute of Technology, Kanpur, 208016, India
\\
$^2$National Center for Radio Astrophysics, Post Bag 3, Ganeshkhind, Pune, 
411007, India}

\date{March 2020}

\begin{document}
\label{firstpage}
\pagerange{\pageref{firstpage}--\pageref{lastpage}}

\maketitle
\begin{abstract}
    We study the alignment of radio galaxies axes using the FIRST catalogue. we impose several cuts in order to select the candidates which are most likely to be free of systematic bias. In our study we primarily focus on testing for alignment among sources within a certain angular separation from one another since for most sources redshift information is not available. We find a very significant effect for angular distances less than 1 degrees. The distance scale of alignment is found to be roughly 28 Mpc, in agreement with earlier estimates, assuming that these sources are dominantly at redshift of 0.8. However, we are not able to entirely rule out the possibility of systematic bias in data. We also perform a full three dimensional analysis using a smaller data sample for which redshift information is available. In this case we only find a very weak signal at much larger distances.

\end{abstract}

\section{Introduction}
There exist several observations which indicate a correlation between structures at large distance scales. For example, the quasar polarizations at optical frequencies show alignment with one another at very large distance scales, of order Gpc \citep{Hutsemekers:1998,Hutsemekers:2000fv,Jain:2003sg,Hutsemekers:2005iz}. 
The alignment effect is seen to be particularly strong in the direction of the Virgo supercluster \citep{Hutsemekers:1998,Hutsemekers:2000fv}.
A similar phenomenon has also been observed in the case of radio polarizations from distant sources. In this case
the correlation was seen on a smaller scale of about 100 Mpc \citep{Tiwari:2013pol,Pelgrims:2015,Tiwari:2016}. Furthermore radio jets from distant galaxies were found
to be aligned on a distance scales greater than 20 Mpc \citep{Taylor:2016rsd}. The signal has been further tested by \cite{2017MNRAS.472..636C}. This paper uses the Radio Galaxy Zoo data and finds a significant signal at small distance scales in the range [19,38] Mpc $h^{-1}_{70}$.  \cite{2020A&A...635A.102B} 
use Very Large Baseline Interferometry (VLBI) maps from Astrogeo database and do not find a significant signal at distance scales larger than 60.5 Mpc which is consistent with the results  obtained earlier.
We point out that the radio polarizations and jet axis are in general correlated
with one another, although at some relatively high frequencies 
this correlation is absent \citep{2020A&A...635A.102B,Tiwari:2018yfb}. Hence, excluding some high frequency observations, we expect that if either one of these show the alignment effect, it should be present in the other observable also.
The effect in both optical and radio data needs to be tested further in order to confirm its presence. Ideally there should be a dedicated study for this purpose so that the systematic effects can be controlled effectively. 

Theoretically, there have been large number of proposals aimed at explaining the alignment effect both at optical and radio frequencies. Some of these are based on intrinsic properties of the galaxies and other invoke propagation effects. 
The radio alignment, which is seen dominantly at small distance scales \citep{Tiwari:2013pol,Pelgrims:2015,Taylor:2016rsd,2017MNRAS.472..636C}, may be explained by invoking correlated magnetic fields within a cluster
of galaxies \citep{Tiwari:2016}. If this is indeed applicable then it will also most likely imply alignment of optical polarizations at such scales; however, a detailed study of such a connection has so far not been done. The observed optical alignment is seen on much larger distance
  scales \citep{Hutsemekers:2014} and may be explained theoretically by invoking 
hypothetical light pseudoscalar particles \citep{Jain:2002vx,Agarwal:2009ic,Piotrovich:2009zz,Agarwal:2012}, vector perturbations \citep{Morales:2007rd}, dark energy coupled to magnetic field
\citep{Urban:2009sw}, cosmic strings \citep{Poltis:2010yu,Hackmann:2010ir}, 
anisotropic expansion \citep{Ciarcelluti:2012pc}
and a superhorizon mode in the Universe \citep{Chakrabarty:2016}. Many of these effects, such as the mixing of electromagnetic waves with hypothetical pseudoscalars \citep{Jain:2002vx,Agarwal:2009ic,Piotrovich:2009zz,Agarwal:2012}, are frequency dependent and would not apply at radio frequencies. 
The recent observation of alignment of galaxies within poor clusters \citep{2020arXiv200608835T} is consistent with the theoretical model proposed in \cite{Tiwari:2016}. The alignment arises due to the cosmological magnetic field which may originate in the early Universe \citep{Turner:1987bw,Ratra:1991bn,Dolgov:1993vg,Gasperini:1995dh,Subramanian:2015lua}. As the sources evolve and interact with field galaxies the alignment may disappear \citep{2020arXiv200608835T}.

In the current paper, we study the alignment of source position angles (PA) using the VLA FIRST catalogue \citep{1995ApJ...450..559B,1997ApJ...475..479W,2015ApJ...801...26H}. The basic idea is to test if the axes of different sources within a certain angular distance
are aligned with one another. 
The FIRST (Faint Images of the Radio Sky at Twenty one cm) survey is a large area sky survey covering about 80 percent of the North galactic cap and 20 percent of the south galactic cap. The total sky area is about 10,000 square degrees. The survey has evolved from its first catalog \citep{1995ApJ...450..559B} to the latest catalog released in 2014 \citep{2015ApJ...801...26H}. It uses images centered at frequencies 1365 and 1435 MHz. Over most of the area the threshold of detection is 1 mJy. 
By design, the FIRST survey area overlaps with that of the Sloan Digital Sky Survey (SDSS)
 and roughly 40 \% of the sources have optical counterparts. The data is processed and self-calibrated by an automated pipeline. 
 The redshifts of sources are not available but the median redshift is known to be about 0.8 \citep{1989ApJ...338...13C}. Although there is significant overlap with SDSS, due to the high redshifts of radio sources it is difficult to get a significant population of sources even with photometric redshifts. Hence we shall restrict most of our study to  alignment of sources within a certain angular separation. We also obtain redshifts of as many sources as possible using the catalog compiled by \cite{2008AJ....136..684K}. This catalog contains radio sources from four radio surveys, i.e. FIRST, NVSS, WENSS, and GB6 with their redshifts obtained from the optical SDSS survey. For these sources we perform a full three dimensional analysis.  However, the number of sources in this study as well as their redshift range is limited and should be repeated with a larger sample. 

Due to the large sky coverage and considerable overlap with optical survey, the FIRST catalog is well suited for our study. It has a high density of sources which allows test of alignment at small angular scales. This would allow us to further study the alignment effect seen in radio polarizations \citep{Tiwari:2013pol,Pelgrims:2015,Tiwari:2016} and radio axes \citep{Taylor:2016rsd}. 
We should point out that our study is somewhat different from that performed in \cite{Taylor:2016rsd}. Here we shall focus primarily on the position angles of individual sources. In contrast, \cite{Taylor:2016rsd} study the alignment among axes joining two lobes of a radio galaxy. The latter study can also be conducted using the FIRST catalog but we postpone it to future research. The FIRST catalog also has
 wide sky coverage and hence allows a test at large angular scales in order to determine if an alignment effect at distance scales seen in  optical frequencies \citep{Hutsemekers:1998,Hutsemekers:2000fv,Jain:2003sg,Hutsemekers:2005iz} is also present in radio. However, the latter test is best performed in three dimensions. Here we undertake this study with a limited data set for which redshifts are available through the SDSS survey \citep{2008AJ....136..684K}. As mentioned earlier, this study should be repeated in future once more detailed redshift information becomes available. 
 
 The FIRST survey has also been used in an earlier study \citep{2017MNRAS.472..636C}. However, our sample selection is considerably different. We rely entirely on the FIRST survey whereas \cite{2017MNRAS.472..636C} uses the Radio Galaxy Zoo data. Furthermore we impose a limit of minor axes $b_{min}$ such that $b_{min}>7$ arcsecs on our data sample, whereas \cite{2017MNRAS.472..636C} impose $b_{min}>2$ arcsecs.  As we discuss below some of the other cuts also differ from ours. We find that our results differ from them in detail but qualitatively we find agreement.

\section{Data}
\label{sec:data}
 The FIRST data set contains a total of 946432 sources \citep{1995ApJ...450..559B,1997ApJ...475..479W,2015ApJ...801...26H}. A large number of sources are very close to one another. These are likely to be lobes of the same radio galaxy. Including such sources will lead to spurious signal of alignment since lobes of the same galaxy are likely to be aligned with one another. We remove these sources by eliminating  all sources within one arcmin of one another.  With this cut we select only those sources which do not have any other source within one arcmin of itself. After imposing this cut we obtain a total of 666021 sources. In \cite{2017MNRAS.472..636C}, such sources, which are
 very close to one other, are handled by passing a line through the sources and the slope of the resulting line is taken as the position angle.
We also remove all sources for which the minor axis ($b_{min}$),  obtained after deconvolution in the catalogue, is smaller than 7 arcsec. This is imposed since the beam itself is about 5.4 arcsec FWHM in the Northern 
hemisphere and a small beam
asymmetry may introduce a bias in the PA.  This cut eliminates a very large number of the sources, with only 8893 sources remaining after this cut. In the Southern hemisphere, declination less than 4 degrees 33 min  21 sec, the beam is elliptical, $6.4\times 5.4$ arcsec FWHM. We find that the sources in Southern hemisphere show a very large signal of alignment which is not consistent with the signal seen in the Northern hemisphere. This is discussed in more detail below. This alignment can be attributed to ellipticity of the beam and hence we only keep sources with Declination greater than 4.6 degrees. After this cut the number of sources are 6955.

\begin{figure}
     \centering
     \includegraphics[width=0.48\textwidth]{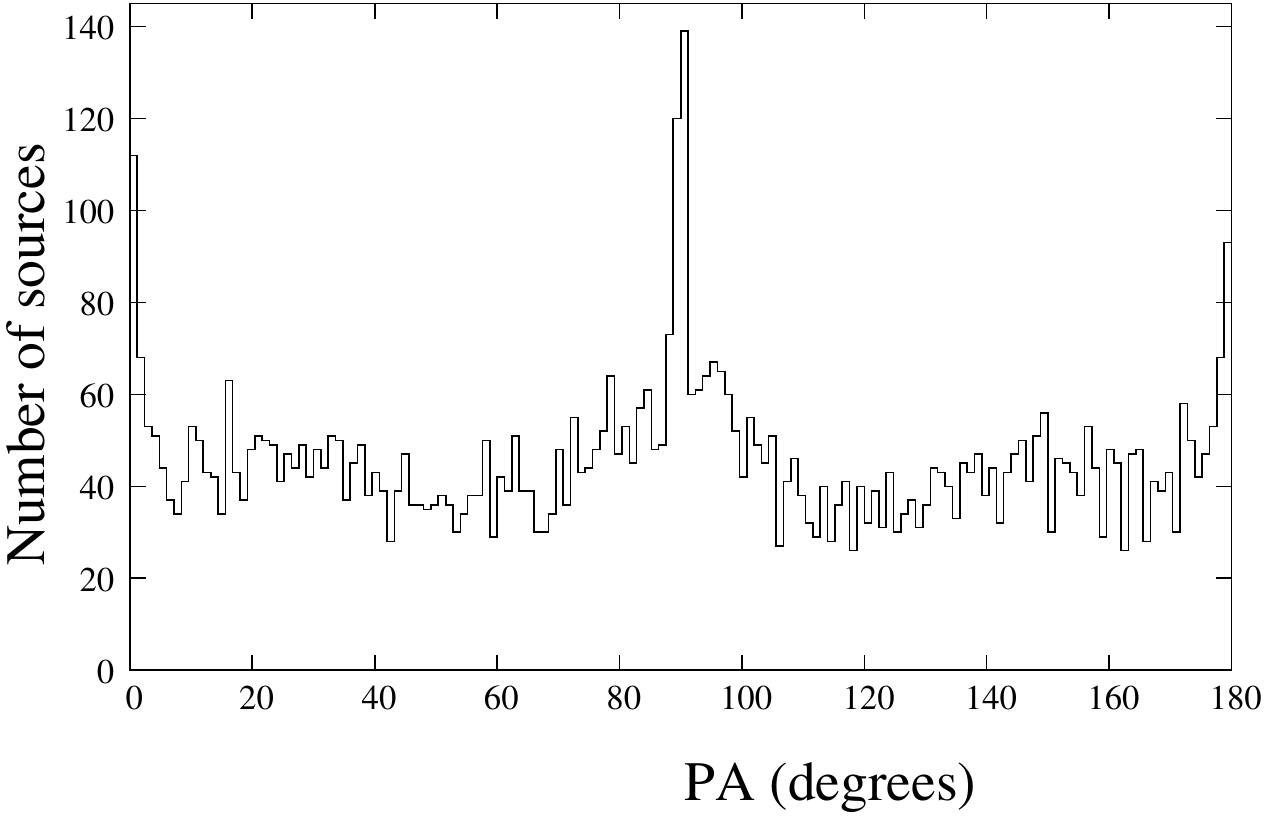}\\
     \caption{ The histogram of PAs after removing all sources with $b_{min}\le 7$ arcsec, along with other cuts (source angular separation greater than 1 arcmin and $Dec > 4.6$ degrees) described in text. }
     \label{fig:distcut0}
\end{figure}

\begin{figure}
     \centering
     \includegraphics[width=0.48\textwidth]{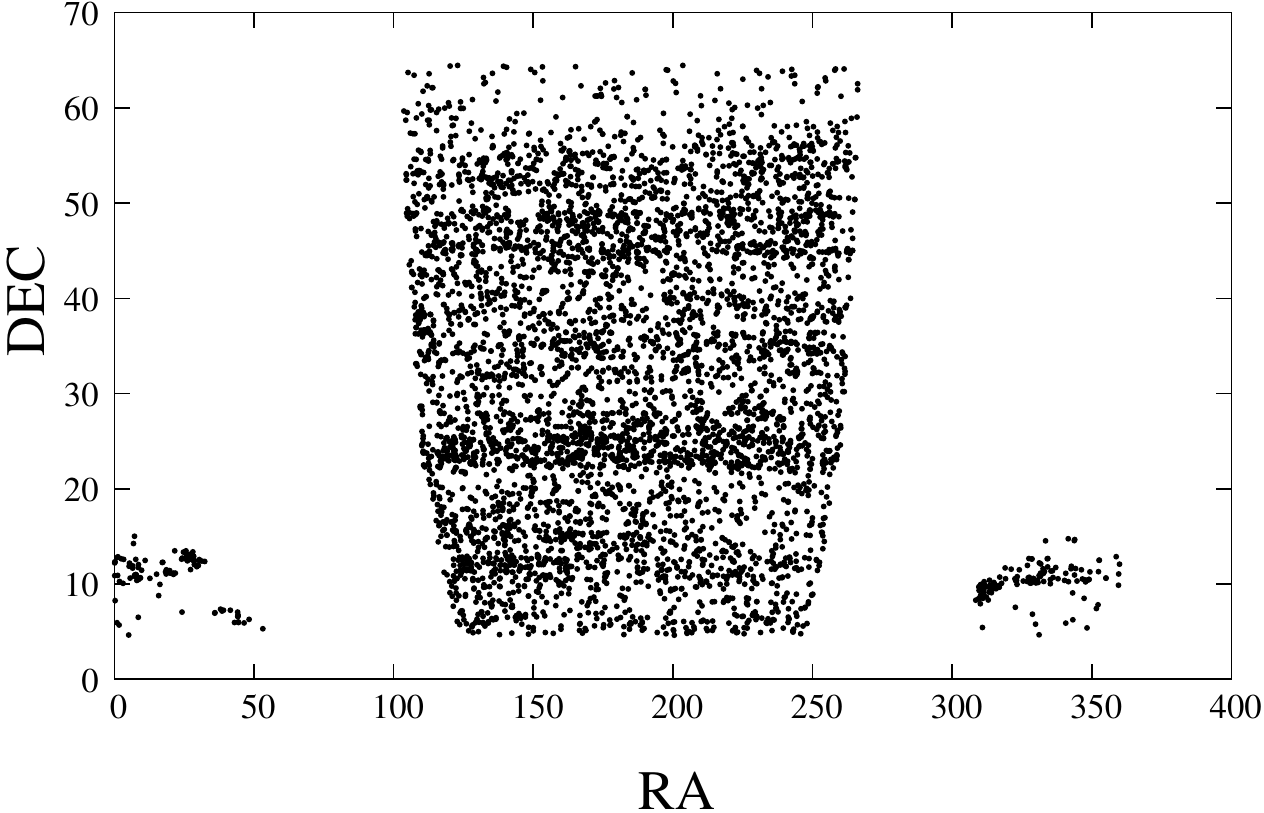}\\
     \caption{The scatter plot of sources after imposing cut 1.     }
     \label{fig:catalog}
\end{figure}

\begin{figure}
     \centering
     \includegraphics[width=0.48\textwidth]{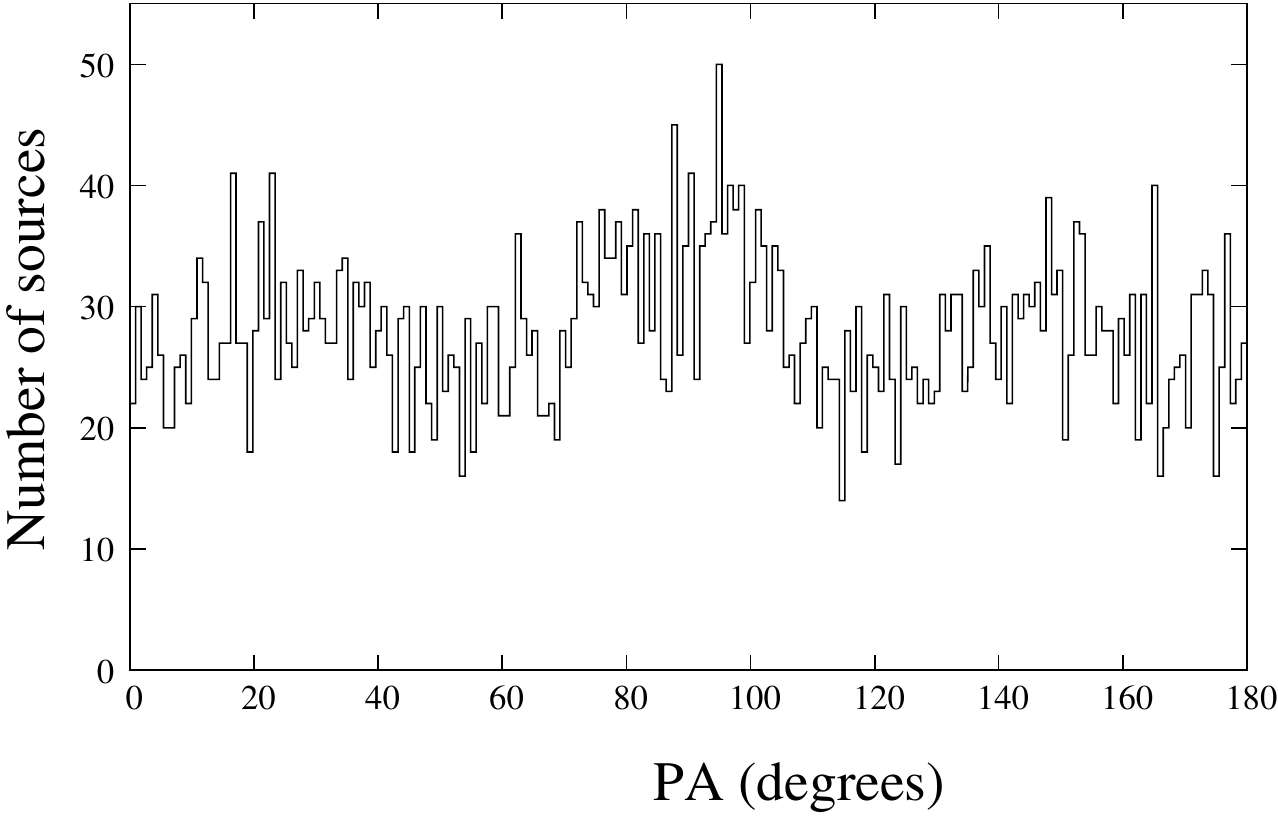}\\
     \caption{ The histogram of PAs after imposing cut 1. This involves removing all sources with $b_{min}\le 7$ arcsec and $(b_{maj} - b_{min})/b_{maj}) > 0.12$, along with other cuts described in text.    }
     \label{fig:distcut8}
\end{figure}

The histogram of PAs of all the sources after imposing the cuts described above is shown in Fig. \ref{fig:distcut0}. We see that the distribution shows sharp peaks at PA=0, 90 and 180 degrees. These are spurious and arise due to sources for which the difference between major ($b_{maj}$) and minor axes is very small. Hence we require that the ratio $(b_{maj} - b_{min})/b_{maj}) > 0.12$. We find that as we make this cut more stringent the spurious peaks at PA=0, 90 and 180 degrees get reduced eventually getting saturating once the ratio takes the value 0.12. This justifies our use of the precise upper limit on this ratio. The number of sources remaining after this cut, which we refer to as cut 1, are 5855. The scatter plot of sources is shown in Fig. \ref{fig:catalog}. The histogram of PAs after this cut is shown in 
Fig. \ref{fig:distcut8}. We still see fluctuations in this distribution, however the sharp peaks at PA=0, 90 and 180 degrees are eliminated. We point out that for angular data there is no analogue of the central limit theorem and hence no preferred distribution, such as a normal distribution. However, in the present case we would have expected the distribution to be uniform. Instead we find a prominent peak at 90 degrees and less prominent peaks roughly at PA =25 and 150 degrees. The reason for these peaks is not clear. This might indicate some large scale correlation in data. Here we shall not focus on this effect and our statistical test will use random samples which will preserve this distribution. Hence the signal we test for is independent of the nature of this distribution.

The scatter plot in Fig. \ref{fig:catalog} shows that the data has some disjoint regions which contain relatively small number of sources. As discussed later, by direct calculation we find that these small regions lead to spurious results for large angular separations. Hence it is best to remove these regions by imposing a cut $RA>75$ degrees and $RA<300$ degrees. 
We will refer to this as cut 2 and use this data set for further studies of alignment. The number of sources remaining after this cut is 5619. The histogram of PAs with this cut remains similar to that with cut 1, shown in Fig. \ref{fig:distcut8}.

\begin{figure}
     \centering
     \includegraphics[width=0.48\textwidth]{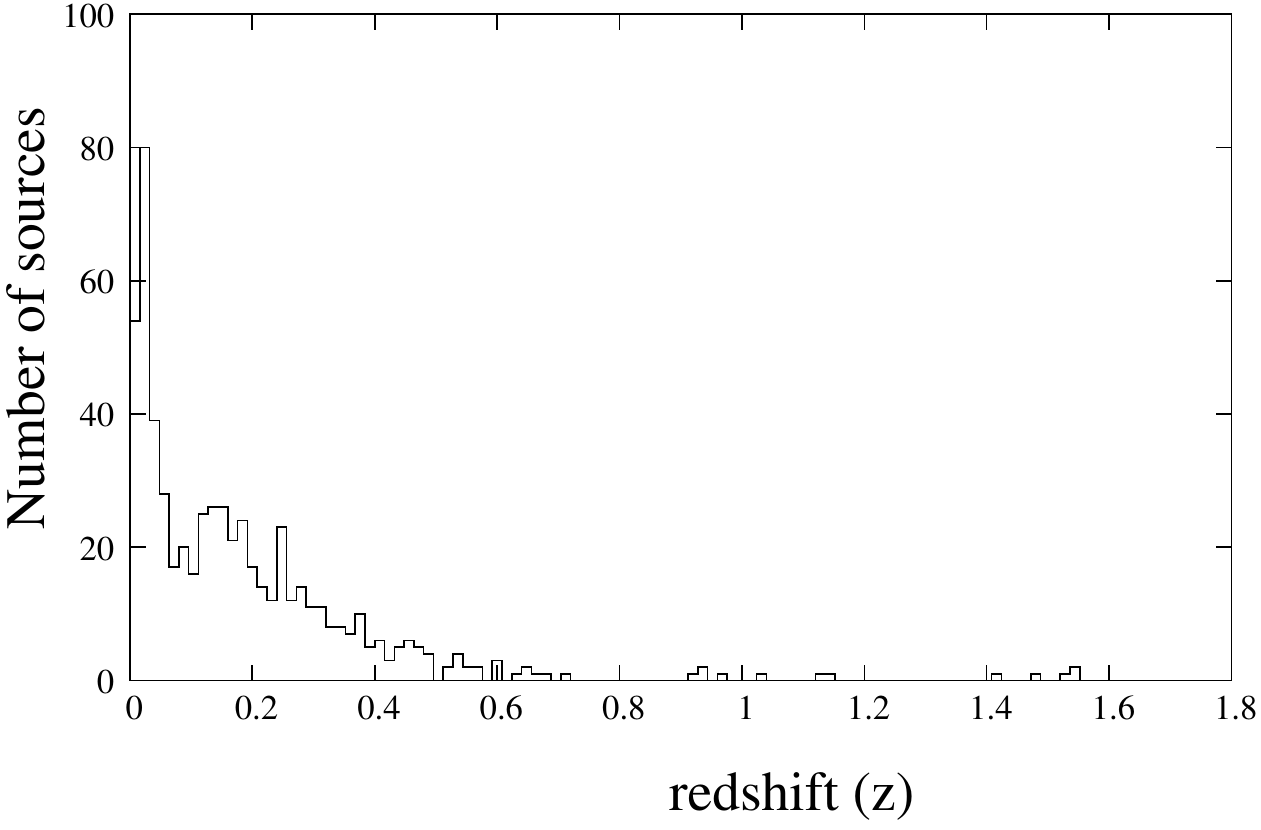}\\
     \caption{ The histogram of redshifts after imposing cut 2. The total number of sources for which redshifts are available after imposing this cut at 593.    }
     \label{fig:zdist}
\end{figure}

In the cuts imposed so far we removed all sources which are within 1 arcmin of one another. Instead of removing such sources we can also test for alignment by imposing a minimum angular separation of 1 arcmin. The analysis procedure is described in more detail below. In this case we do not need to impose the 1 arcmin cut but impose all the other cuts corresponding to cut 2. These cuts are, $b_{min}> 7$ arcsec, $Dec>4.6$ degrees, $(b_{maj} - b_{min})/b_{maj}) > 0.12$, $RA> 75$ degrees and $RA<300$ degrees. We refer to this as cut 3 and the total number of sources remaining after this cut are 18775.

For our three dimensional analysis we use the combined catalog \citep{2008AJ....136..684K} which uses radio data from FIRST, NVSS, WENSS and GB6 and optical data from SDSS. After all the cuts described above, corresponding to cut 2 (or cut 3), we find a total of 593 sources which have redshift information. The redshift distribution of this set is shown in Fig. \ref{fig:zdist}. We see that the distribution is sharply peaked at low $z$ with very few sources having $z$ greater than 0.5.
Hence it will only allow a limited test of alignment. It would have been interesting to test the alignment seen in optical sources at high redshifts \citep{Hutsemekers:1998,Hutsemekers:2000fv}. 
However, the redshift range of the present data is much smaller and hence it is not possible to reliably test the alignment on the distance scale seen in optical sources.

\section{Procedure}
We follow the procedure outlined in \citep{Hutsemekers:1998,Jain:2003sg} in order to test for alignment. Let $\psi_i$ be the position angle of the i$^{th}$ source. For each source position angle $\psi_i$ we define a vector $\vec v_i$ on the surface of the celestial sphere with components $[\cos(2\psi_i),\sin(2\psi_i)]$. In order to compare these vectors for two sources $i$ and $j$ on the celestial sphere we need to parallel transport the vector at site $i$ to site $j$. A convenient measure of alignment of two sources i and j is given by
\begin{equation}
    d_{ij} = \cos[(2\psi_i+\Delta_{i\rightarrow j})-2\psi_j]
\end{equation}
where the factor $\Delta_{i\rightarrow j}$ arises due to the parallel transport. This factor is negligible for sources at small angular separation and can be ignored in this case. However, for large angular separations it cannot be neglected. 

In most of our study we shall be testing for alignment of sources within a certain angular separation since we have limited redshift information and cannot compute the distance of most of the sources. Consider the $k^{th}$ source. Let there be $n_k$ sources within an angular separation $\Delta \beta$ of this source. Here we shall include the $k^{th}$ source also in the $n_k$ nearest neighbours although this will have no effect on our results. A measure of alignment of this source with other sources is given by
\begin{equation}
    D_k(\Delta\beta) = \frac{1}{n_k}\sum_{i=1}^{n_k} d_{ik}\Bigg|_{\theta<\Delta\beta}
    \label{eq:Dk}
\end{equation}
where $\theta$ is the angular separation between two sources.
Let $N$ be the total number of sources in the data set. We define the statistic
\begin{equation}
    S_D = \frac{1}{N} \sum_{k=1}^N D_k
\end{equation}
as a measure of alignment over the entire sample. 

In order to determine the significance of the signal we generate random samples by randomly permuting the PAs among different sources while keeping the source positions fixed. This is useful in order to preserve the distribution of PAs. We generate a large number of random samples and determine the resulting distribution of the statistic $S_D$ which can then be compared with the data value. As we shall see the significance in many cases is found to be relatively large. So a direct numerical estimate of significance is impractical. Instead we determine the mean and standard deviation of the statistics in the random samples and use these values to compute the significance. We use 25000 random samples for computing the significance in most cases.

It is also useful to get 
an alternate measure of alignment which gives an effective angular spread of PAs in data. We consider all sources within an angular separation $\Delta\beta$ of the $k^{th}$. Let $\psi_{k}^{av}$ be the mean angle of this set. It is obtained by maximizing the following measure $A_k$ of alignment:  
\begin{equation}
    A_k = \frac{1}{n_k}\sum_{i=1}^{n_k} \cos[(2\psi_i+\Delta_{i\rightarrow k})-2\psi_k^{av}]\Bigg|_{max}
    \label{eq:Ak}
\end{equation}
The average value $A$ of $A_k$ over the entire sample provides a useful measure of alignment and $\cos^{-1}A$ provides an effective measure of angular spread of $2\times$PAs in data. 
We point out that while averaging over $A_k$ we do not include those sources which have no neighbours within the angular spread $\Delta\beta$.

The above procedure can be generalized so that we only consider sources which are at angular separation larger than a minimum value $\Delta\beta_0$ in Eq. \ref{eq:Dk}. This is useful for the case of cut 3 in which we do not want to include sources which are within 1 arcmin of one another. We therefore generalize Eq. \ref{eq:Dk} as
\begin{equation}
    D_k(\Delta\beta,\Delta\beta_0) = \frac{1}{n_k}\sum_{i=1}^{n_k} d_{ik}\Bigg|^{\theta<\Delta\beta}_{\theta>\Delta\beta_0}
    \label{eq:Dk1}
\end{equation}
where for most of our analysis we shall set $\Delta\beta_0 = 1$ arcmin. The rest of the procedure remains the same as before.

For our three dimensional analysis we use the statistic defined in Eq. \ref{eq:Dk} with the nearest neighbours selected by using the comoving distance as a measure of separation among sources. We use the $\Lambda CDM$ model with $\Omega_\Lambda=0.69$ and $\Omega_M=0.31$.

\section{Results}
\begin{figure}
     \centering
     \includegraphics[width=0.48\textwidth]{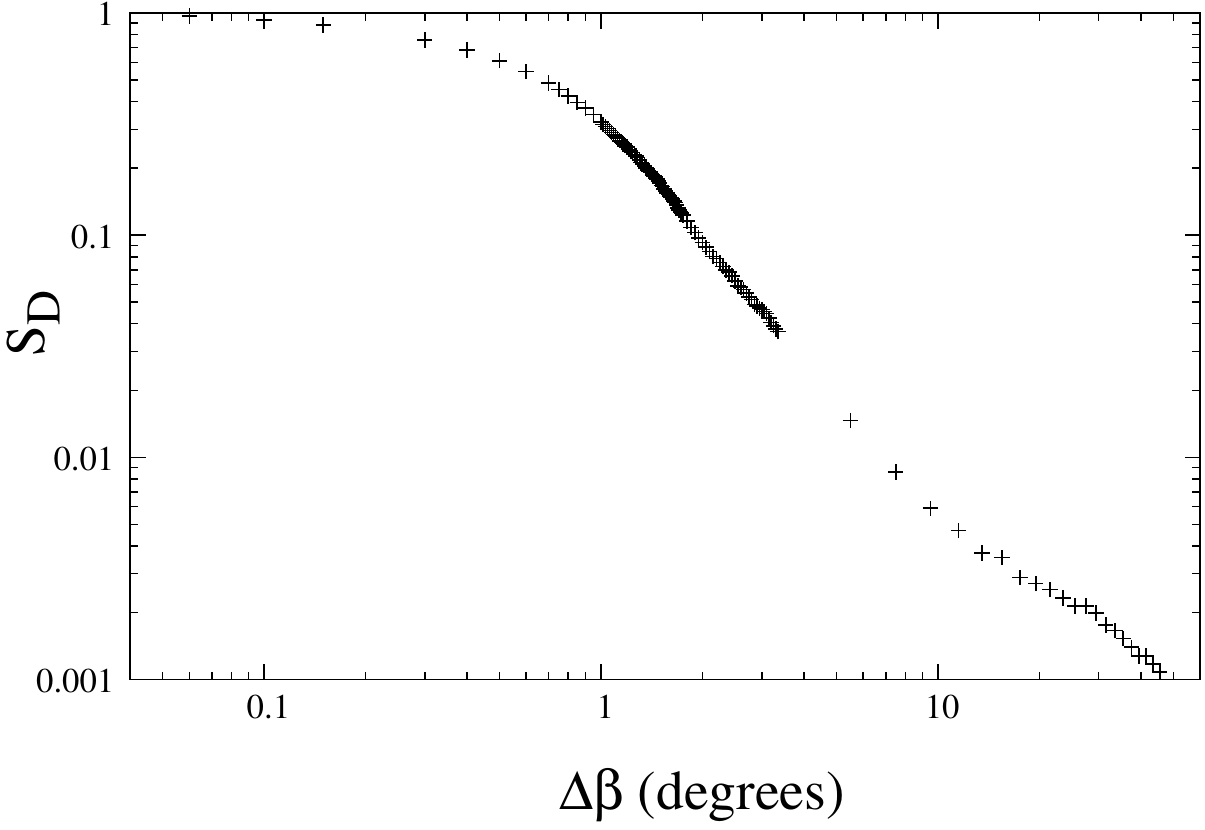}\\
     \caption{The statistic $S_D$ as a function of the angular separation $\Delta \beta$   }
     \label{fig:SD}
\end{figure}

\begin{figure}
     \centering
     \includegraphics[width=0.48\textwidth]{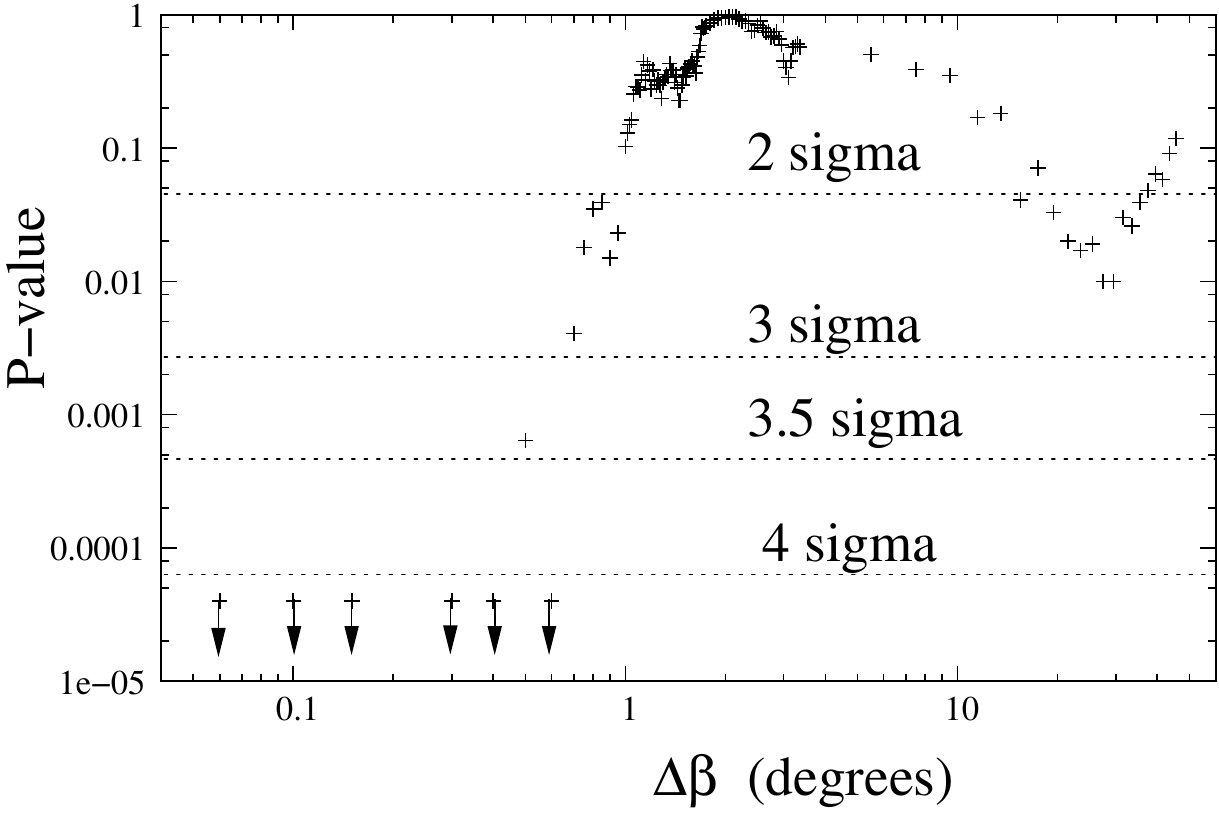}\\
     \caption{The probability that the statistic observed in data can arise as a random fluctuation as a function of the angular separation $\Delta \beta$. The sigma levels are indicated for convenience. We have obtained the P-values by generating a maximum of 25000 random samples. The lower 6 points indicate that the P-value is smaller than 1/25000.  }
     \label{fig:sigma}
\end{figure}

The statistic $S_D$ for cut 2 as a function of the angular separation 
 $\Delta\beta$ is shown in Fig. \ref{fig:SD}. The corresponding P-values, that is the probability that the statistic can arise as a random fluctuation, are shown in Fig. \ref{fig:sigma}. The sigma values are also shown on this plot.  We find that $S_D$ decreases rather sharply for small values of $\Delta\beta$ and starts to decay gently for larger values, $\Delta\beta>10$ degrees. In Fig. \ref{fig:sigma} we see that for small values of $\Delta\beta$, less than 1 degree, the significance is very large (small P-values), nearly 4 sigmas. We have computed the P-values by generating a maximum of 25000 random samples. In some cases this did not generate even a single random sample with statistic larger than the data value. Hence in these cases the P-value is smaller than $4.0\times 10^{-5}$ and is indicated by downward pointing arrows in Fig. \ref{fig:sigma}. We find that the significance is very large, consistently more than 4 sigmas, for $\Delta\beta \le 0.7$ degrees.
 The significance reduces with increase in $\Delta\beta$ and we do not observe a significant  alignment effect for $\Delta\beta>1.5$ degrees. For larger values of $\Delta\beta$ the P-values start to become smaller for $\Delta\beta$ in the neighbourhood of 25 degrees and reach about 2.5-sigma significance level. 
 
 We point out that if did not impose the cut on declination, the statistical significance is much higher for all values of  $\Delta\beta$. This indicates a clear difference between data in the Northern and the Southern hemisphere which suggests presence of a systematic effect, justifying our cut on Declination. The problem is associated with the ellipticity of the beam, as described in section \ref{sec:data}. Furthermore if we did not remove the small disjoint regions by imposing the cut on RA, we find high significance for large values of $\Delta\beta$ but negligible change for small values of this variable.

We next determine how the results may change if instead of drawing the random samples from the data distribution, we drew them from a uniform distribution. For this purpose we simply compare the mean and standard deviations of the $S_D$ values obtained from a large number of random samples generated from the two different distributions of PA. For this test we set $\Delta\beta=1$ degree. For the random samples obtained from the data distribution, Fig. \ref{fig:distcut8} we obtain the mean and standard deviation to be 0.2093 and $3.6\times 10^{-3}$ respectively. The corresponding values for a uniform distribution are found to be 0.2088 and $3.6\times 10^{-3}$ respectively. Hence we find negligible change for $\Delta\beta=1$ degree. For large angular separations
$\Delta\beta=45$ degree we do find a significant difference, as expected. In case for the data distribution we find mean and standard deviation to be $5.11\times 10^{-4}$ and $1.29\times 10^{-4}$ respectively and for uniform distribution, $1.20\times 10^{-4}$ and $1.23\times 10^{-4}$ respectively. The two are clearly different. We also find that for this case the data statistic value is  $5.78\times 10^{-4}$. It is clear that we do not get a significant signal of alignment if we compare this value with random samples generated from the data distribution. However, we do get a significant effect if we use a uniform distribution. A direct numerical calculation shows that the significance is a little less than 3 sigmas (P-value $\approx 0.01$), although the mean and standard deviation values may suggest a higher significance. This difference arises since the distribution differs from a Gaussian for large values of the statistic. Hence we find that although the non-uniform nature of the PA distribution in Fig. \ref{fig:distcut8} suggests a large scale correlation the effect is not very significant.

We next determine the quantity $A$, i.e. average value of $A_k$ defined in Eq. \ref{eq:Ak}. For this calculation we set $\Delta\beta=0.6$ degree. This value is chosen since beyond this the significance drops significantly. We obtain $A=0.533$. This corresponds to an angular spread $2\times$PA of 58 degrees. Hence the PAs appear to be aligned within an angular region of $\pm 29$ degrees. 

\begin{figure}
     \centering
     \includegraphics[width=0.48\textwidth]{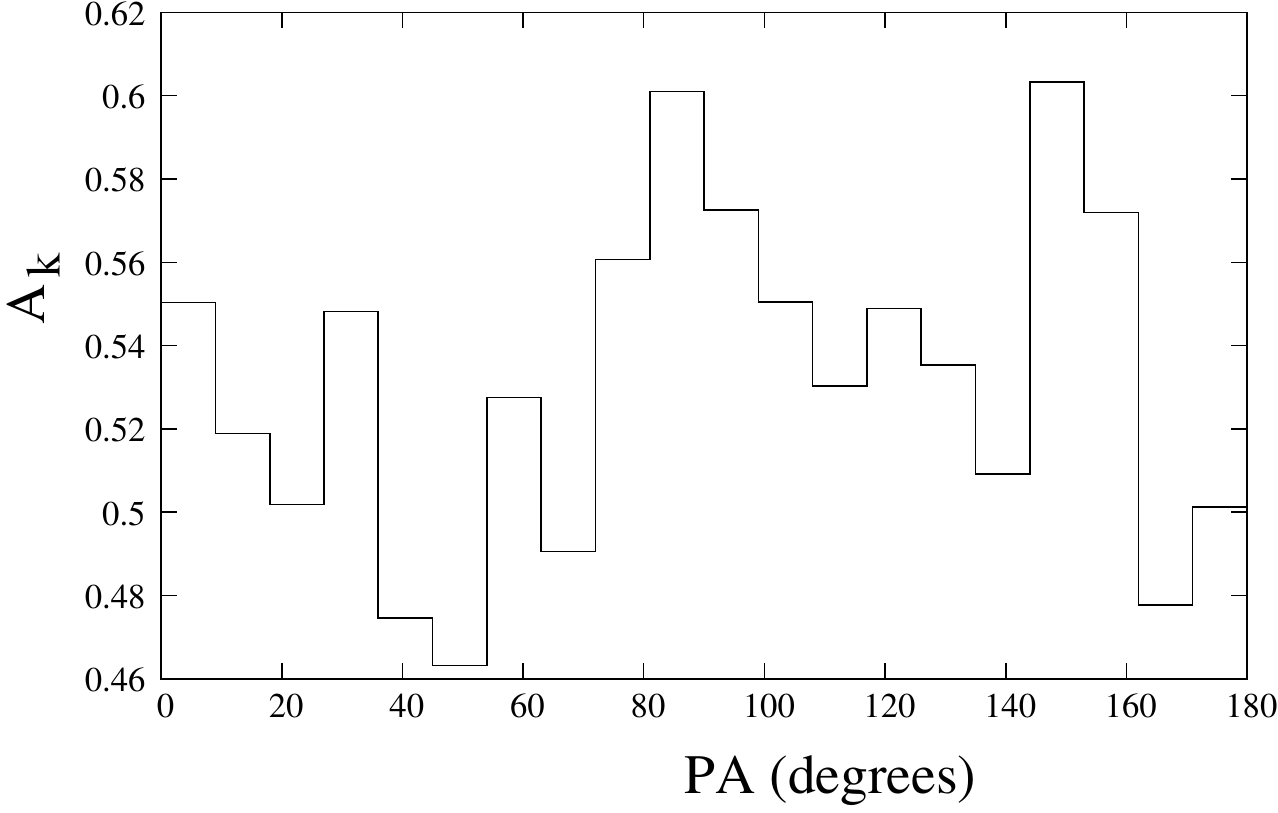}\\
     \caption{The binned measure of alignment $A_k$ (Eq. \ref{eq:Ak}) as a function of PA for $\Delta\beta = 0.6$ degree.  }
     \label{fig:alignment}
\end{figure}

We next determine which values of PAs give dominant contribution to alignment for the case of $\Delta\beta=0.6$ degree. In Fig. \ref{fig:alignment} we plot binned values of $A_k$ as a function of PA. We find a peak around PA=85 degrees and another around PA=145 degrees.

 In Fig. \ref{fig:sigma} we see that the alignment effect is seen dominantly on angular scales of less than 0.7 degrees. This scale is similar to the size of the co-added images. On such scales it is possible to have alignment if there are significant
beam and/or sidelobe effects present in data. In order to study this possibility we performed two additional tests:
\begin{itemize}
    \item[1.] We investigate if there are 
any sources present in our sample for which the PA before ($\text {PA}_\text {before}$) and after ($\text {PA}_\text {after} $) deconvolution differs by a large amount. Such sources are likely to have significant contributions from beam asymmetry effects and can lead to a spurious signal. We specifically imposed the cut $|\text {PA}_\text {before} - \text {PA}_\text {after}| <15$ degrees. We find that all of the sources in our sample satisfy
this cut and no additional sources get eliminated. 
\item[2.] We compute the error in the major and minor axes of all the sources in order to determine if the sources are well resolved. We use the following empirical formula to estimate the error $\sigma$ in size \citep{1997ApJ...475..479W},
\begin{equation}
    \sigma = 10{''}\ \left({1\over {\text SNR}} + {1\over 75}\right) \,.
\end{equation}
We find that the ratio $b_{maj}/\sigma >4$ and $b_{min}/\sigma > 3.2$ for all the sources in our final data sample. This implies that all the sources considered in our study are well resolved. 
\end{itemize}
This analysis gives us further 
confidence that the data sample is reliable and the signal we observe may be caused by a physical effect.

\subsection{Results with Cut 3}
\begin{figure}
     \centering
     \includegraphics[width=0.48\textwidth]{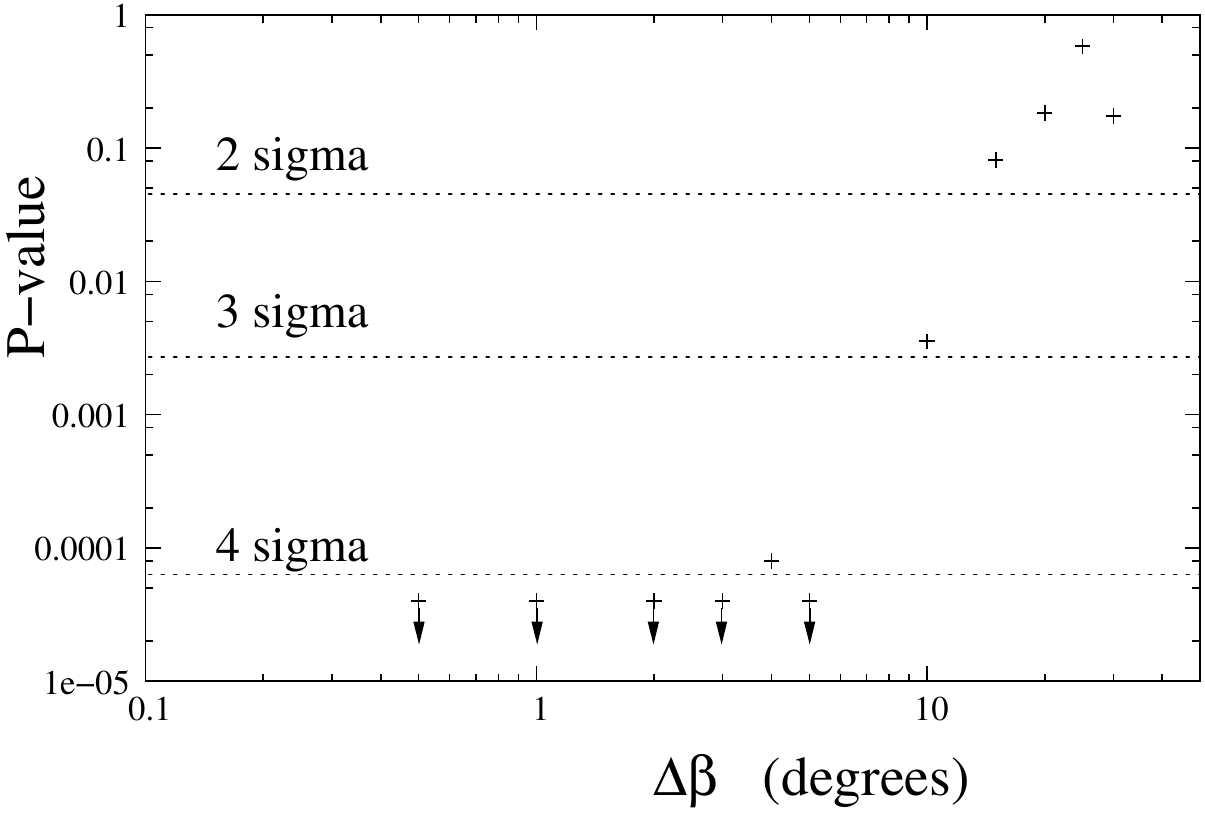}\\
     \caption{The P-values for the case of cut 3 as a function of the upper limit on $\Delta\beta$ with the lower limit set at 1 arcmin. The P-values are generated using 25000 random samples. The lowest data points indicate that no random sample exceeded the data statistic. }
     \label{fig:rescut3}
\end{figure}

\begin{figure}
     \centering
     \includegraphics[width=0.48\textwidth]{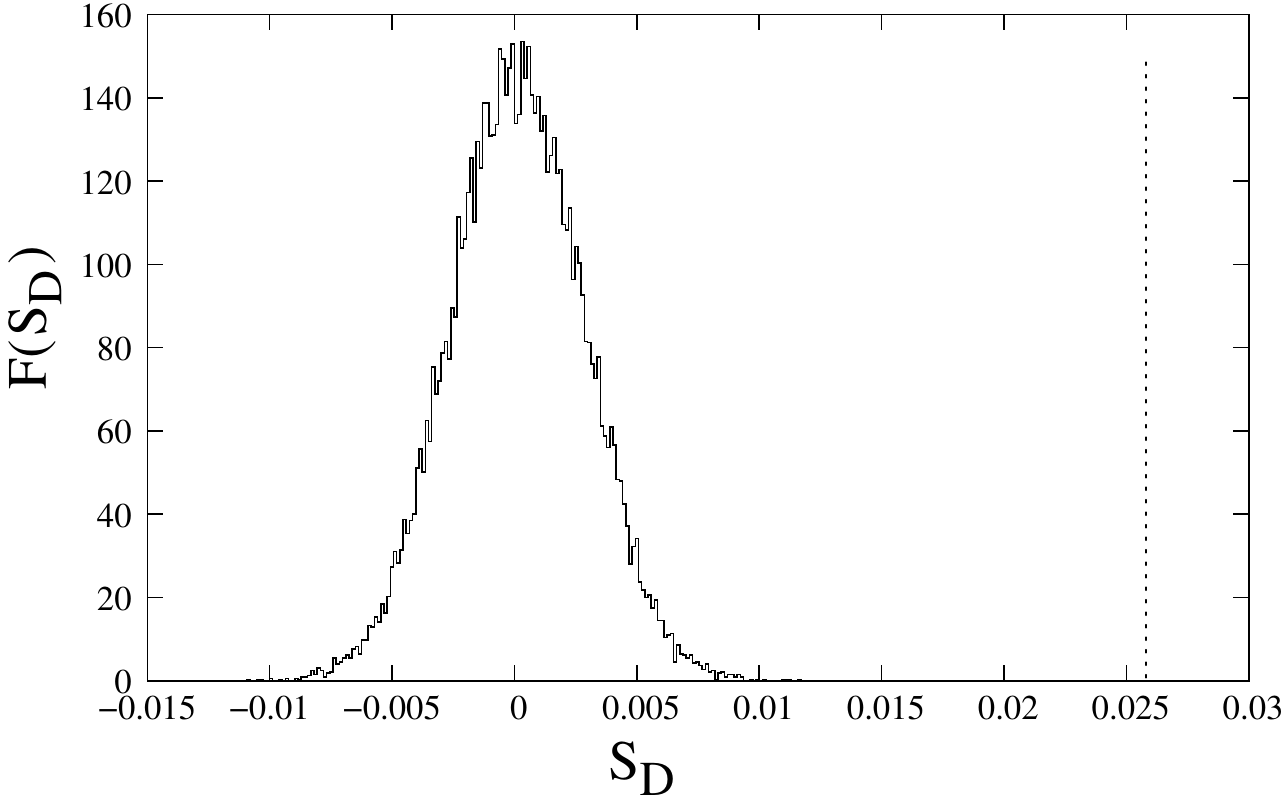}\\
     \caption{The normalized distribution of statistic $S_D$ for random samples for the upper limit on $\Delta\beta=1$ degree. The dashed line at the right shows the data statistic.  }
     \label{fig:hist1deg}
\end{figure}

In this case we have a larger data sample of 18775 sources and we test for alignment only for sources which have angular separation greater than 1 arcmin. Hence we use Eq. \ref{eq:Dk1} in computing our statistic. We find that this sample shows a much stronger signal of alignment for small angular separations. The results are shown in Fig. \ref{fig:rescut3}. We see a very strong signal of alignment up to angular separations $\Delta \beta$ of 5 degrees. For larger angular separations, greater than 15 degrees, we do not see a significant signal. For small separations the signal is actually much stronger in comparison to what is indicated in Fig. \ref{fig:rescut3}, which is based on using 25000 random samples. We show the distribution of $S_D$ for the random samples in Fig. \ref{fig:hist1deg} along with the data statistic for the upper limit on $\Delta\beta=1$ degree. The mean and standard deviation of the distribution is found to be $1.3\times 10^{-4}$ and $2.72\times 10^{-3}$ respectively. With the data statistic value of $2.58\times 10^{-2}$, this implies a significance of 9.4 sigmas. We find that the significance decreases monotonically as we increase the angular separation. 

It is also useful to test in which range of angular separations we get significant alignment. It is very clear from our results that small separations lead to a very strong signal. However, it is possible that larger separations are only adding noise and the signal is dominantly arising from very small separations. In order to determine the regions which gives significant contribution we repeat the test of alignment in different regions. In particular we test for the region between $\Delta\beta_0 = 0.75$ degree and $\Delta\beta = 1.25 $ degree. We find that in this region the significance of alignment is approximately 2.5 sigmas and hence is not negligible. For angular intervals centered at larger values the $\Delta\beta$ the significance drops considerably. Hence we conclude that the alignment is present only on angular scales up to about 1 degree.

Finally we determine the distance scale of alignment within the $\Lambda CDM$ model. The dominant signal of alignment is seen upto angular distance of 1.0 degrees. The sources in this catalog peak at the redshift ($z$) of about 0.8. Assuming this value we can determine the distance scale of separation corresponding to 1 degree angular separation at $z=0.8$ using $\Lambda CDM $ model with  $\Omega_\Lambda = 0.69$, $\Omega_M=0.31$ and $H_0=67.7$ Km sec$^{-1}$Mpc$^{-1}$ and spatially flat metric. The relationship between the angular diameter $\Delta\beta$ and physical length $L$ of the object at redshift $z$ is given by
\begin{equation}
    L = d_A \Delta\beta
\end{equation}
where $d_A$ is the angular diameter distance. 
This leads to a distance scale $L$ of approximately 28 Mpc which is in agreement with earlier result on the alignment of galaxy axis \citep{Taylor:2016rsd} but somewhat smaller than that obtained in \cite{Tiwari:2013pol}.

\subsection{Three dimensional analysis}
\begin{figure}
     \centering
     \includegraphics[width=0.48\textwidth]{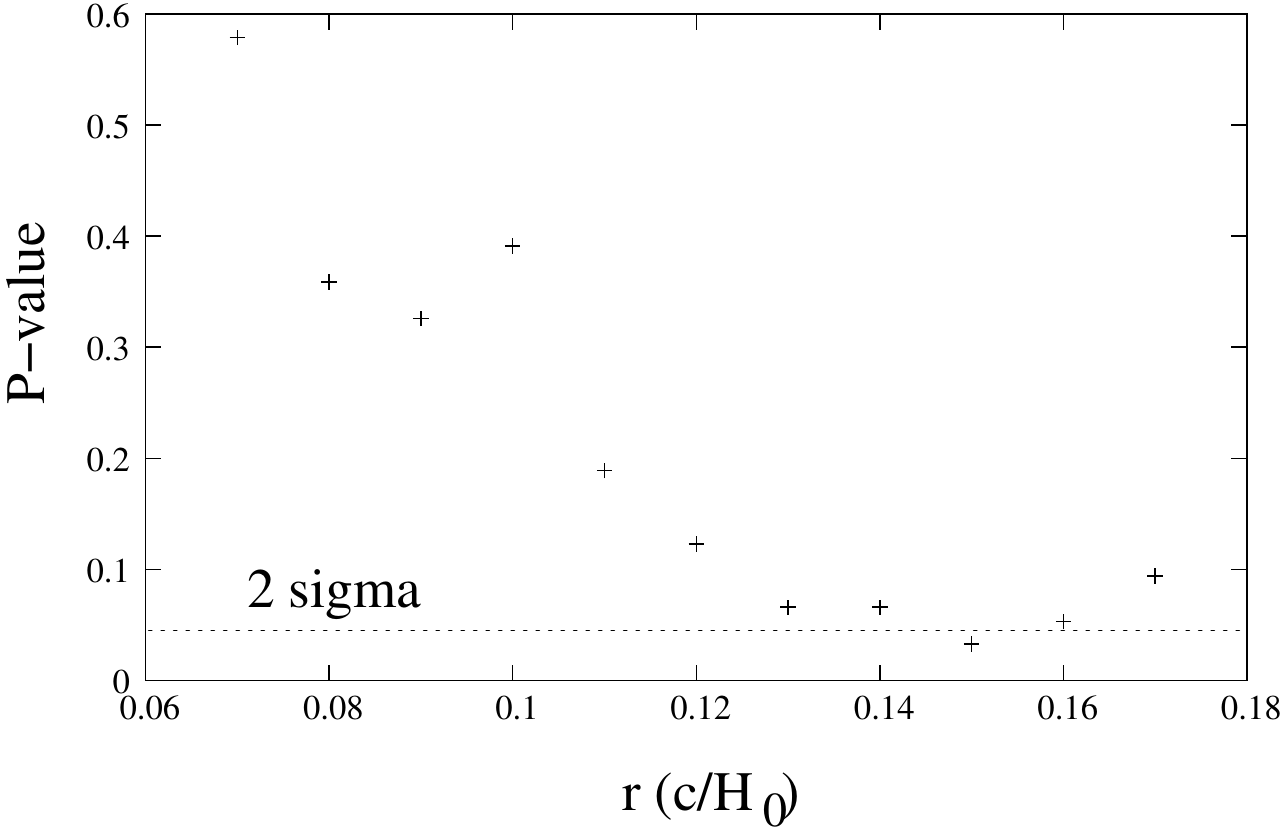}\\
     \caption{The probability that the statistic observed in data can arise as a random fluctuation as a function of the comoving distance $r$ in units of $c/H_0$. The two sigma levels is indicated for convenience. Here we use the three dimensional analysis with the cut $z\le 0.5$.   }
     \label{fig:sig3d}
\end{figure}

In this section we present the results of the three dimensional analysis using 593 sources for which redshifts $z$ are available from the SDSS survey \citep{2008AJ....136..684K}. In this case we use the comoving distance $r$ as a measure of separation. We use the $\Lambda CDM$ model with $\Omega_\Lambda = 0.69$ and $\Omega_M=0.31$ and spatially flat metric. We test for alignment over a range $r=0.05$ to $r=0.17$ in units of $c/H_0$. For the full sample we do not find a significant signal for any of these values. We also try a cut such that $z\le 0.5$, which is motivated by the fact that there are very few sources with $z$ is greater than this value. Hence these would act as outliers. With this cut the $P-values$ are shown in Fig. \ref{fig:sig3d}. We find a very mild signal for $r\approx 0.15 c/H_0$ which corresponds to a distance of approximately 640 Mpc. 

\section{Conclusions}
We have studied the alignment of radio axes using the FIRST catalog. We impose several cuts on the data in order to select a sample which may be free of bias arising due to asymmetric beam or due to irregular sky coverage. In most of our analysis we restrict our work to a test of alignment among sources projected on the celestial sphere and separated within a certain maximum angle. This is because for most of these sources redshift information is not available. We also perform a full three dimensional test for a subset of sources for which the redshift information is available. For our two dimensional test, we found that the radio sources show a strong signal of alignment over angular scales of order 1 degree or smaller. This corresponds to a distance scale of 28 Mpc within $\Lambda CDM$ model, consistent with earlier observations  \citep{Taylor:2016rsd}. We also see a mild effect at larger angular scales of order 25 degrees in one of our  samples. Here the significance is a little larger than 2 sigmas. However, this effect is absent in a larger data sample and hence is likely to be a statistical fluctuation. The effect seen at smaller angular scales is consistent with what has been seen earlier. However, in order to firmly establish it we need to rule out the possibility of systematic bias in data. In our analysis we have identified several sources of bias and imposed several cuts in order to reduce such effects. However, the possibility that the effect could arise from beam effect cannot be entirely ruled out since the dominant alignment is seen only over small angular scales. A dedicated study to extract the position angles relatively free of systematic effects is needed to order to firmly establish the result. 

In our three dimensional analysis, which uses a significantly reduced data sample, we do not find a very significant effect. In this case due to limited number of source we are unable to test at very small distances and hence the absence of effect is consistent with our results of the two dimensional test. After eliminating a few outliers at $z>0.5$ we do find a very mild effect at distance scale of 640 Mpc. This may be tested further with enhanced data sample.

\bigskip
\noindent
{\bf Acknowledgements:} The authors acknowledge funding from the Science and Engineering Research Board (SERB), Government of India, grant number EMR/2016/004070 for this research work. 

\bigskip
\noindent
{\bf Data Availability:} No new data were generated or analysed in support of this research.

\bibliographystyle{mnras}
\bibliography{master}

\bsp
\label{lastpage}
\end{document}